\documentclass[sigconf]{acmart} 
\pdfoutput=1
\usepackage[shortlabels]{enumitem}
\usepackage{balance}
\usepackage{dblfloatfix}

\usepackage{hyperref}
\usepackage{cleveref}

\makeatletter
\let\th@plain\relax
\makeatother

\newcommand{\bi}{\begin{itemize}[leftmargin=0.4cm]}
	\newcommand{\ei}{\end{itemize}}
\newcommand{\be}{\begin{enumerate}[leftmargin=0.4cm]}
	\newcommand{\ee}{\end{enumerate}}

\definecolor{Gray}{gray}{0.85}
\usepackage{tikz}
\usepackage{framed}
\usepackage[framed]{ntheorem}
\usetikzlibrary{shadows}

\theoremclass{Lesson}
\theoremstyle{break}

\tikzstyle{thmbox} = [rectangle, rounded corners, draw=black, fill=Gray!40]

\newshadedtheorem{lesson}{Result}

\begin{document}

\copyrightyear{2018} 
\acmYear{2018} 
\setcopyright{acmcopyright}
\acmConference[ICSE-SEIP '18]{40th International Conference on Software Engineering: Software Engineering in Practice Track}{May 27-June 3, 2018}{Gothenburg, Sweden}
\acmBooktitle{ICSE-SEIP '18: 40th International Conference on Software Engineering: Software Engineering in Practice Track, May 27-June 3, 2018, Gothenburg, Sweden}
\acmPrice{15.00}
\acmDOI{10.1145/3183519.3183549}
\acmISBN{978-1-4503-5659-6/18/05}

\title{We Don't Need Another Hero?}
\subtitle{The Impact of ``Heroes''  
on  Software Development }

\author{Amritanshu Agrawal, Akond Rahman, Rahul Krishna, Alexander Sobran* and Tim Menzies}
\affiliation{Computer Science, NCSU, USA; IBM Corp*, Research Triangle, North Carolina}  
\email{[aagrawa8, aarahman, rkrish11]@ncsu.edu, asobran@us.ibm.com, tim@menzies.us}

\begin{abstract}
A  software project has 
  ``Hero  Developers'' when  80\% of contributions are delivered by 20\% of the developers. Are such heroes a good idea? Are too many heroes bad for  software quality?  Is it better to have more/less heroes
for different kinds of projects?
To answer these questions, we studied 661 open source  projects from Public open source software (OSS) Github and 171  projects from an Enterprise Github.

We find that hero projects
are very common. In fact, as projects grow in size, nearly
all projects become hero projects. These   findings  motivated us to look more closely at the
effects of heroes on software development.
Analysis shows that the frequency to close issues and bugs are not significantly
affected by the presence of heroes or project type (Public or Enterprise). Similarly, 
the time needed to resolve an issue/bug/enhancement is not affected by heroes or project type. This is a surprising result since, before looking at the data, we expected that increasing
heroes on a project will slow down how fast that project reacts to change. 
However, we do find a statistically significant association between heroes,  project types, and enhancement resolution rates. Heroes do not affect enhancement resolution rates in  Public projects. However, in Enterprise projects,
heroes increase the rate at which projects complete enhancements.

In summary, our empirical results call for a revision of a long-held truism in software engineering.
Software heroes are far more common and valuable than suggested by the literature,
particularly for medium to large  Enterprise developments. Organizations should reflect on better ways
to find and retain more of these heroes.

\end{abstract}



\maketitle

\section{Introduction}
Many  projects are initiated by a project leader who stays
in that  project for the longest duration~\cite{ye2003toward}. These leaders are the ones who moderate the projects, contributes the most, and stays the most active throughout  the
software development life cycle. Such developers are sometimes called Hero/ Core/ lone contributors~\cite{martinez2014current}.  
In the literature~\cite{goeminne2011evidence,torres2011analysis,robles2009evolution,yamashita2015revisiting},
it is usual to define a {\em hero project} as one where 80\% of the contributions are made by 20\% of the developers.

In the literature, it is usual to deprecate heroes~\cite{bier2011online,morcovcomplex,hislop2002integrating,boehm2006view,wood2005multiview}
since they can become a bottleneck that slows down project development.
  That said, looking through the literature, we cannot see any   large
scale studies on the effect of heroes in Enterprise projects.
Accordingly, to better understand the positive or negative impact of heroes in software development, we  mined 661 Public open source software (OSS) projects  and 171 Enterprise Github projects
(we say that  enterprise software are in-house proprietary projects that used public Github Enterprise repositories to manage their development).
After applying    statistical tests to this data, we found some surprises:
\bi
\item Hero projects are exceedingly  common  in both Public and Enterprise  projects, and the ratio of hero programmers in a project does not affect the development process, at least for the metrics we looked, with two exceptions;
\item Exception \#1: in larger projects,  heroes are far  {\em more} common, that is, large projects need their heroes;
\item Exception \#2:  heroes have a {\em positive impact} on Enterprise projects, specifically,
the {\em more heroes}, the {\em faster the enhancement resolution rates}  to those kinds of projects. 
\ei
This was surprising since, before mining the data, our expectation was that heroes have a large negative effect on software
development, particularly for Public projects where the work is meant to be spread around a large community.

The rest of this paper explains how we made and justified these findings. This investigation is
structured around the following research questions:
\bi
    \item \textbf{RQ1: How common are heroes?}
    
     From this analysis, we found:
     
\ei
   
 \begin{lesson}  Over 77\%  projects exhibit the pattern that 20\% of the
    total contributors complete 80\% of the contributions. This holds
    true for both  Public and Enterprise projects .
\end{lesson}

\bi
    \item \textbf{RQ2: How does team size affect the prevalence of hero projects?}
    
    After dividing teams into  small, medium and large sizes,  we found that:
    
\ei
    \begin{lesson} As team size increased, an increasing proportion of projects become
    hero projects. This is true for both Public and  Enterprise projects.
    \end{lesson}

\bi
    \item \textbf{RQ3:  Are hero projects associated with  better software quality?}
    
    We extracted 6 quality measures, namely number of issues, bugs and enhancements being resolved, and the time taken to resolve these issues, bugs and enhancements.
    
    \bi
        \item \textbf{a: Does having a hero programmer improves the number of issues, bugs and enhancements being resolved?}
    \ei
        \begin{lesson} For both Public and Enterprise projects, there is no statistical difference between the percent of
       issues and bugs being resolved within  hero and non-hero projects.
        However, for  enhancement issues, Enterprise/Pubic hero projects closed statistically more/less issues (respectively).
        \end{lesson}
        
    \bi
        \item \textbf{b: Does having a hero programmer improves the time to resolve issues, bugs and enhancements?}
    \ei
        \begin{lesson} There was no statistically difference found in the resolution times of issues, bugs and enhancements among non-hero and hero projects in either cases (Public or Enterprise).
        \end{lesson}
\ei

Based on the above, 
we say that our empirical results call for a revision of a long-held truism in software engineering.
Software heroes are far more common and valuable than suggested by the literature,
particularly for medium to large  Enterprise developments. Organizations should reflect on better ways
to find and retain more of these heroes.

The rest of this paper is structured as follows.
Following this introduction,  Section \ref{sec:literature} gives a literature review regarding Hero programmers in OSS then
Section~\ref{sec:experiment} describes   the data extraction process and the experimentation details.
The  research questions are answered in
Section~\ref{sec:results} and the implications of these results are discussed in Section~\ref{sec:discuss}. Finally, we discuss the validity and conclusion of our results.

\section{Background and Related Work}
\label{sec:literature}

\subsection{Project Roles}
\label{sec:hero}

Following on from Ye and Martinez et al.~\cite{ye2003toward,martinez2014current}, we say that
there  are many developer roles within a Public or Enterprise software project:
\bi
\item
 Project leaders,  who initiate a project;
 \item
 Core members, who work on the project and make many contributions over an extended
 time periods;
 \item
 Active developers, who contributes regularly for new enhancements and bug fixes;
 \item
 Peripheral developers, who occasionally contributes to new enhancement;
 \item
 Bug fixers;
 \item
 Bug reporters;
 \item
 Bug readers;
 \item
 Passive users.
 \ei
Of the above,  Core developers can be project leaders or core members. 
Core developers are the few central developers who implement most of the code
changes and make important project direction decisions, whereas  the other peripheral developers being the ``many eyes'' of the project that make small changes such as bug fixes~\cite{mockus2002two,tsay2014influence}

Core developers are said to contribute roughly 80\% of the code
who are just about 20\% of their project team size~\cite{goeminne2011evidence,torres2011analysis,robles2009evolution,yamashita2015revisiting}. These contributions can be recorded in terms of how many commits they made or how many lines of code (loc) they changed. Research studies~\cite{krishnamurthy2002cave, peterson2013github}  suggested that most work/contributions are done by lone developer. A core committer is also the one who has write access to a
project's repository~\cite{padhye2014study}. These developers are also called  Hero Programmers.

Pinto et al.~\cite{pinto2016more} studied 275 OSS projects and found that about 48\% of the developers population committed only 1.73\% of the total number of commits (which we are calling peripheral developers). Even in these contributions, about 28.6\% contributions are done simply to fix typos, grammar and issues, 30.2\% tried fixing bugs, 8.9\%  contributions were to refactor code and while only 18.7\% was used to contribute for new features. Yamashita et al.~\cite{yamashita2015revisiting} also found different proportions of contribution activity among the core and peripheral developers.

Since the work in projects is not evenly divided, this motivates our research on the overall
effects on the projects of different levels of contributions by different developers.


\subsection{Related Work}
\label{sec:related}

To the best of our knowledge, the research of this paper is    the largest study on the effects of heroes in 
Public   and Enterprise projects. The rest of this section describes some of the other related
work we have found in this area but it should be noted that none of the following studies
(a)~explore as many projects as we do and (b)~compare effects across Public and Enterprise projects.

The benefits and drawbacks of heroes are widely discussed in the literature.
Bach~\cite{bach1995enough} notes that such heroes are enlisted to  (e.g.,) speed the delivery of late projects~\cite{cullom2006software}.   
On the other hand, hero-based projects have their drawbacks. In hero
projects, there is less collaboration between team members since there
are few active team members. Such collaborations can be highly beneficial.
Studies that analyzed the distributed software development on social coding platforms like Github and Bitbucket~\cite{dias2016does,cosentino2017systematic} commented on how social collaborations can reduce the cost and efforts of software development without degrading the quality of software. 

Distributed coding effort gives rise to agile community-based programming practices which can in turn have higher customer satisfaction, lower defect rates, and faster development times~\cite{moniruzzaman2013comparative, rastogi2017empirical}. Such practices can lead to increased customer satisfaction when faster development leads to:
\bi
\item
Lowering the issues/bugs/enhancements resolution times~\cite{mockus2002two,jarczyk2014github,bissyande2013got,athanasiou2014test,gupta2014process,reyes2017analyzing};
\item
Increasing the  number of issues/bugs/enhancements being resolved~\cite{jarczyk2014github}.
\ei
More specifically, as to issues related to heroes,
Bier et al.~\cite{bier2011online} warn that as project become more and more complex,  teams should
be communities of experts specialized in niche domains rather than being
lead by ``cowboy programmers'' (a.k.a. heroes)~\cite{morcovcomplex}.  
Such hero programmers are often associated with certain process anti-patterns such as poorly documented
systems (when heroes generate code more than documents about that code~\cite{hislop2002integrating}) or
 all-night hackathons to
hastily patch faulty code to meet deadlines, thus introducing more bugs into the system and decreasing the number of people who understand the whole system~\cite{boehm2006view}.  Also, Wood et al.~\cite{wood2005multiview} caution that heroes are often code-focused but software development needs workers
acting as more than just coders (testers, documentation authors, user-experience analysts).

Our summary of the above is as follows: with only isolated exceptions, most of the literature deprecates heroes
even though the value (or otherwise) of heroes in Enterprise software developments has rarely been investigated.
Accordingly, in this paper, we compare and contrast the effects of heroes in Public {\em and} Enterprise development.


\section{Data and Experimentation}
\label{sec:experiment}

\subsection{Data}
\label{sec:data}

To perform our experiments we used OSS projects from public and Enterprise Github. Of the publicly available projects hosted on public Github, a selected set of projects are marked as ``showcases'', to demonstrate how a project can be developed in certain domain such as game development, and music~\cite{gh:showcase}. 
By selecting these Github projects we can ensure we are using  an interesting and representative set of open source projects. Examples of popular projects included in the Github showcases that we used for our analysis are: Javascript libraries such as `AngularJS'\footnote{https://github.com/angular/angular.js} and `npm'\footnote{https://github.com/npm/npm}, and programming languages such as `Go'\footnote{https://github.com/golang/go}, `Rust'\footnote{https://github.com/rust-lang/rust}, and `Scala'\footnote{https://github.com/scala/scala}.

Not all projects hosted on Github are good for the analysis. Studies done by~\cite{kalliamvakou2014promises,bird2009promises,MunaiahCuration2017}
advice that researchers should filter out the projects which will not be suitable for  analysis. Such unsuitable projects might record only
minimal   development activity, are used only for personal purposes, and not even be related to software development at all. Accordingly,
we apply the following filtering rules.

We started off with 1,108  Public and 538 Enterprise Github projects. Following
the advice of others~\cite{kalliamvakou2014promises}~\cite{bird2009promises}, we pruned  as
follows:

\bi
\item{\textit{Collaboration}: Number of pulls requests are indicative of how many other peripheral developers work on this project. Hence, a project must have at least one pull request.}
\item{\textit{Commits}: The project must contain more than 20 commits.}
\item{\textit{Duration}: The project must contain software development activity of at least 50 weeks.}
\item{\textit{Issues}: The project must contain more than 10 issues.}
\item{\textit{Personal Purpose}: The project must not be used and maintained by one person. The project must have at least eight contributors.}
\item{\textit{Releases}: The project must have at least one release.}
\item{\textit{Software Development}: The project must only be a placeholder for software development source code.}
\ei

\begin{table}[!t]
\small
\centering
\caption{Filtering criteria. Starting with 1108+538  public+enterprise projects, we discard  projects that {\em fail}
any of the LHS tests to arrive at 661+171 projects.}
\begin{tabular}{|r|r|r|}
\hline
Sanity check & \multicolumn{2}{c|}{Discarded project count}                     \\ \cline{2-3} 
                                         & \multicolumn{1}{l|}{Enterprise} & \multicolumn{1}{l|}{Public} \\ \hline
No. of Commits $> 20$                           & 68                               & 96                       \\
No. of Issues  $> 10$                           & 60                               & 89                       \\
Personal purpose (\# programmers $> 8$)  & 47                               & 67                       \\
S\/W development only                    & 9                                & 51                       \\ 
Duration $> 50$ weeks                    & 12                               & 46                       \\
No. of Releases $\textgreater 0$                  & 136                              & 44                       \\
Collaboration (\# Pull requests $> 0$)      & 35                               & 54                      \\
\hline
Projects left after filtering                 & 171                              & 661                       \\ 
\hline
\end{tabular}
\label{tab:sanity}
\vspace{-0.2cm}
\end{table}

 After applying these criteria we obtained 661 open source and 171 proprietary projects. We report how many of the projects passed each sanity check in Table~\ref{tab:sanity}. The projects are discarded when the steps given in Table~\ref{tab:sanity} are applied sequentially, from top to bottom, we are left with 661 open-source and 171 proprietary projects. We used the Github API to extract necessary information from these projects and tested each criteria stated above. Upon completion, we obtained a list of projects from which we extract metrics to answer our research questions. We repeated the procedure for both our Public and Enterprise Github data sources.

\subsection{Metric Extraction}
\label{sec:metric}

To answer our research questions, we extracted the number of commits made by individual developers, and if the number of commits made by 20\% of developers is more than 80\% of the commits, they are classified as Hero Projects and all the others were classified into Non-Hero projects
(these thresholds were selected based on the advice of Yamashita et al~\cite{yamashita2015revisiting}). 

Note that Github allows you to merge the pull requests from external developers and when merged, these contributions gets included in the merger contributor as well. \textit{These merges could introduce more contributions to the Hero Developer so to over-inflate the ``Hero effect'', hence, we did not include those pull merge requests}.

 \begin{figure*}[!b]
  \centering
\includegraphics[width=.5\linewidth]{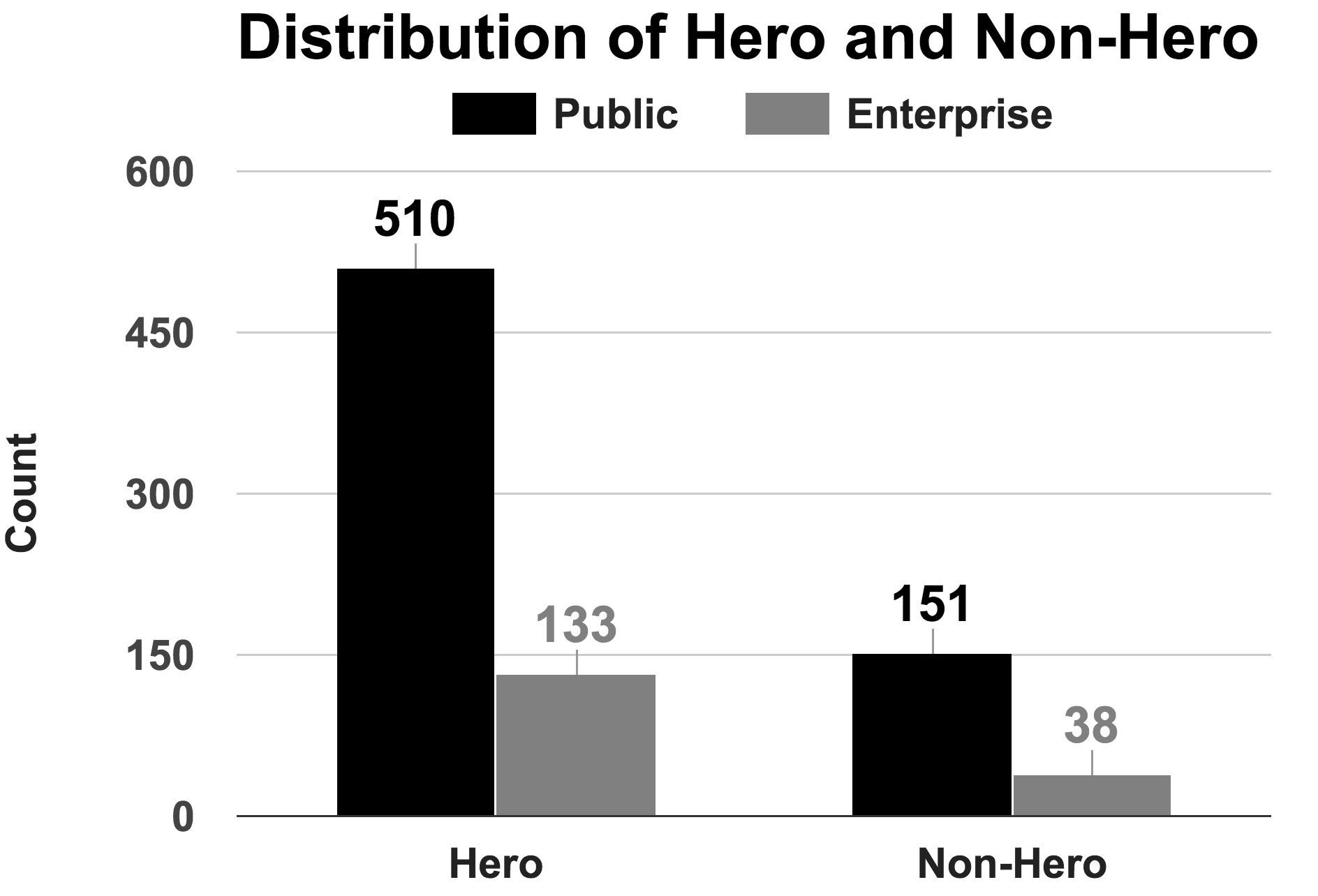}
  \caption{ Distribution of Hero and Non Hero projects in Public and Enterprise projects.
  Note that these hero projects are very common.}
  \label{fig:rq1}
\end{figure*}

We next   divided each project based on the team size. After applying the advice
of Gautam et al.~\cite{gautam2017empirical}, we use 3 team sizes:
\bi
\item {\em Small teams:}  number of developers > 8 but less than 15
\item {\em Medium teams:} number of developers > 15 but less than 30; 
\item {\em Large teams:} number of developers > 30 .
\ei
 
We then defined 6 metrics, namely,
 
{\small
\begin{eqnarray}
I_r\_I_t& =& \frac{\text{Total number of issues closed}}{\text{Total number of issues created}}
    \\
  B_r\_B_t& =& \frac{\text{Total number of Bug tagged issues closed}}{\text{Total number of Bug tagged issues created}}
   \\
     E_r\_E_t& = &\frac{\text{Total number of Enhancement tagged issues closed}}{\text{Total number of Enhancement tagged issues created}}
    \\
     IR_t& = &\text{Median time taken to  resolve   issues}
    \\
    BR_t &=& \text{Median time taken  to resolve   Bug tagged issues}
    \\
   ER_t& =& \text{Median time taken to resolve Enhanced tagged issues}
    \end{eqnarray}}

\subsection{Statistical Tests}
\label{sec:stats}

When comparing the results between Hero and Non-hero, we used a statistical
significance test and an effect size test.
Significance test are useful for detecting if two populations
differ merely by random noise. 
Also, effect sizes are useful for checking that two populations differ by more than just a trivial amount.

For the significance test,  we use the 
     Scott-Knott procedure  recommended at TSE'13~\cite{mittas2013ranking} and ICSE'15~\cite{ghotra2015revisiting}. This
     technique recursively bi-clusters a sorted
    set of numbers. If any two clusters are statistically indistinguishable, Scott-Knott
    reports them both as one group.
    Scott-Knott first looks for a break in the sequence that maximizes the expected
    values in the difference in the means before and after the break.
    More specifically,  it  splits $l$ values into sub-lists $m$ and $n$ in order to maximize the expected value of differences  in the observed performances before and after divisions. For e.g., lists $l,m$ and $n$ of size $ls,ms$ and $ns$ where $l=m\cup n$, Scott-Knott divides the sequence at the break that maximizes:
     \[E(\Delta)=ms/ls*abs(m.\mu - l.\mu)^2 + ns/ls*abs(n.\mu - l.\mu)^2\]
Scott-Knott then applies some statistical hypothesis test $H$ to check if $m$ and $n$ are significantly different. If so, Scott-Knott then recurses on each division.
    For this study, our hypothesis test $H$ was a conjunction of the A12 effect size test (endorsed by
    \cite{arcuri2011practical})  and non-parametric bootstrap sampling \cite{efron94}, i.e., our Scott-Knott divided the data if {\em both}
    bootstrapping and an effect size test agreed that the division was statistically significant (90\% confidence) and not a ``small'' effect ($A12 \ge 0.6$).

\section{Results}
\label{sec:results}

\subsection{RQ1: How common are heroes?}
\label{sec:rq1}

Recall that we define a  project to be heroic when 80\% of the contributions are done by about 20\% of the developers~\cite{yamashita2015revisiting}. To
assess
the prevalence of such projects,
we extracted the above features and classified these projects into hero and non-hero. 

As shown in Figure~\ref{fig:rq1},  77\% and 78\% projects are driven by hero or core developers in Public and Enterprise projects respectively. This trend was also observed by Pinto et al.~\cite{pinto2016more}. 

Why so many heroes?  One explanation is that our results may be incorrect and they are merely a result of the   ``build effect'' reported by Kocaguneli et al.~\cite{ekrem13}.
In their work with Microsoft code files, Kocaguneli et al. initially found an effect that seems similar to heroes. Specifically, in their sample,
most of the files were most often updated by a very small number of developers. It turns out that those ``heroes'' were in fact,   build engineers who had the low-level, almost clerical task of running the build scripts and committed the auto-generated files.
If our results were conflated in the same say then all the results of this paper would be misleading.

We say that our results do not suffer from Kocaguneli build effect, for two reasons:
\bi
\item Kocaguneli reported an extremely small number of build engineers (dozens, out of a total population of thousands of engineers). 
The heroes found in this study
are far more frequent than that.
\item As mentioned before, we did remove any pull merge requests from the commits to remove any extra contributions added to the hero programmer. This means that the contributions aggregated by many developers would not contribute to a few build engineers in our sample.
\ei

\label{sec:rq2}
 \begin{figure*}[!t]
  \centering
\includegraphics[width=.5\linewidth]{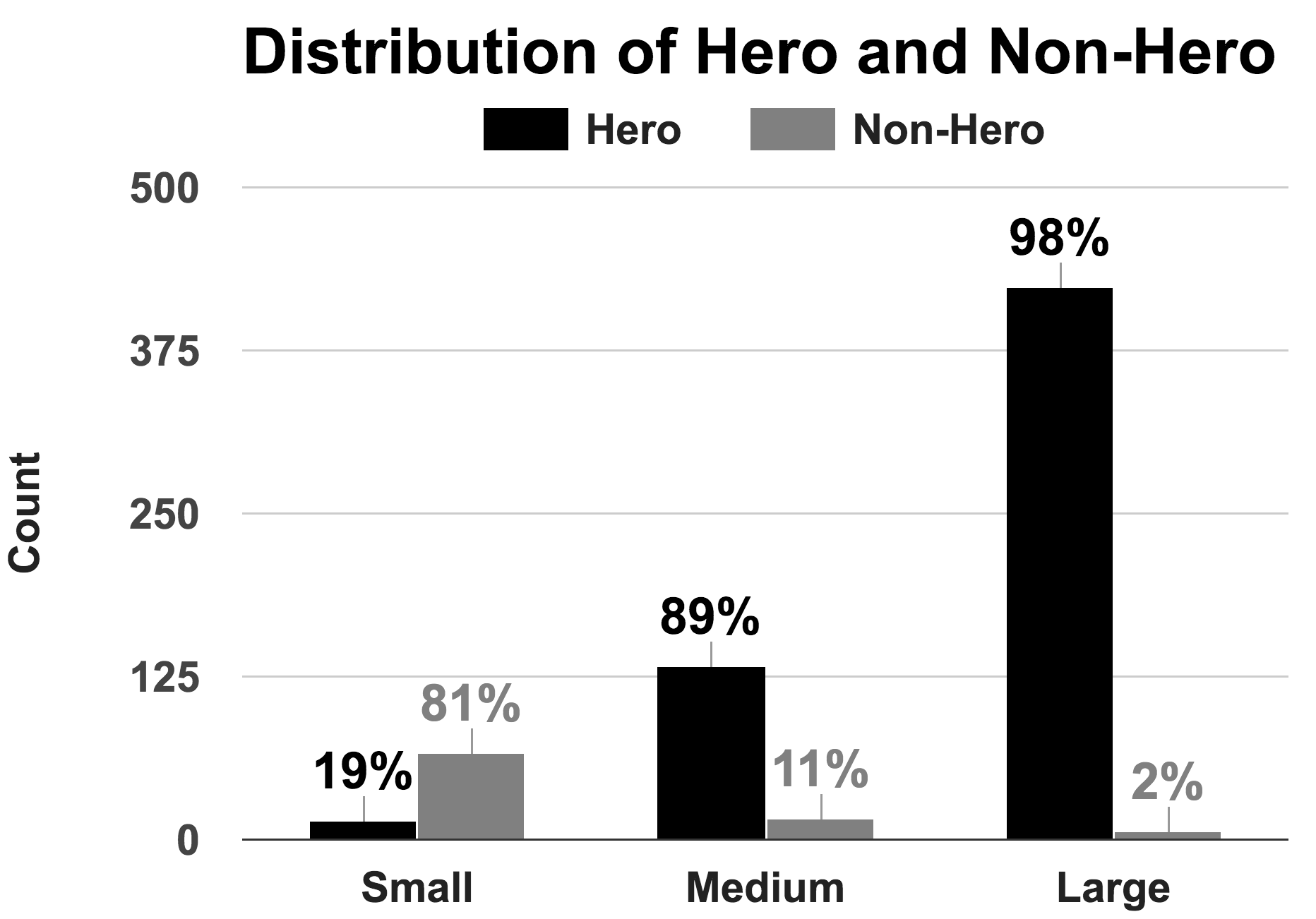}
  \caption{ Public projects:  Hero and Non Hero projects for different team sizes. The percentages shown within the histogram bars show that,  as
  team size grows, the ratio of hero project increases.}
  \label{fig:rq2a}
\end{figure*}

 \begin{figure*}[!t]
  \centering
\includegraphics[width=.5\linewidth]{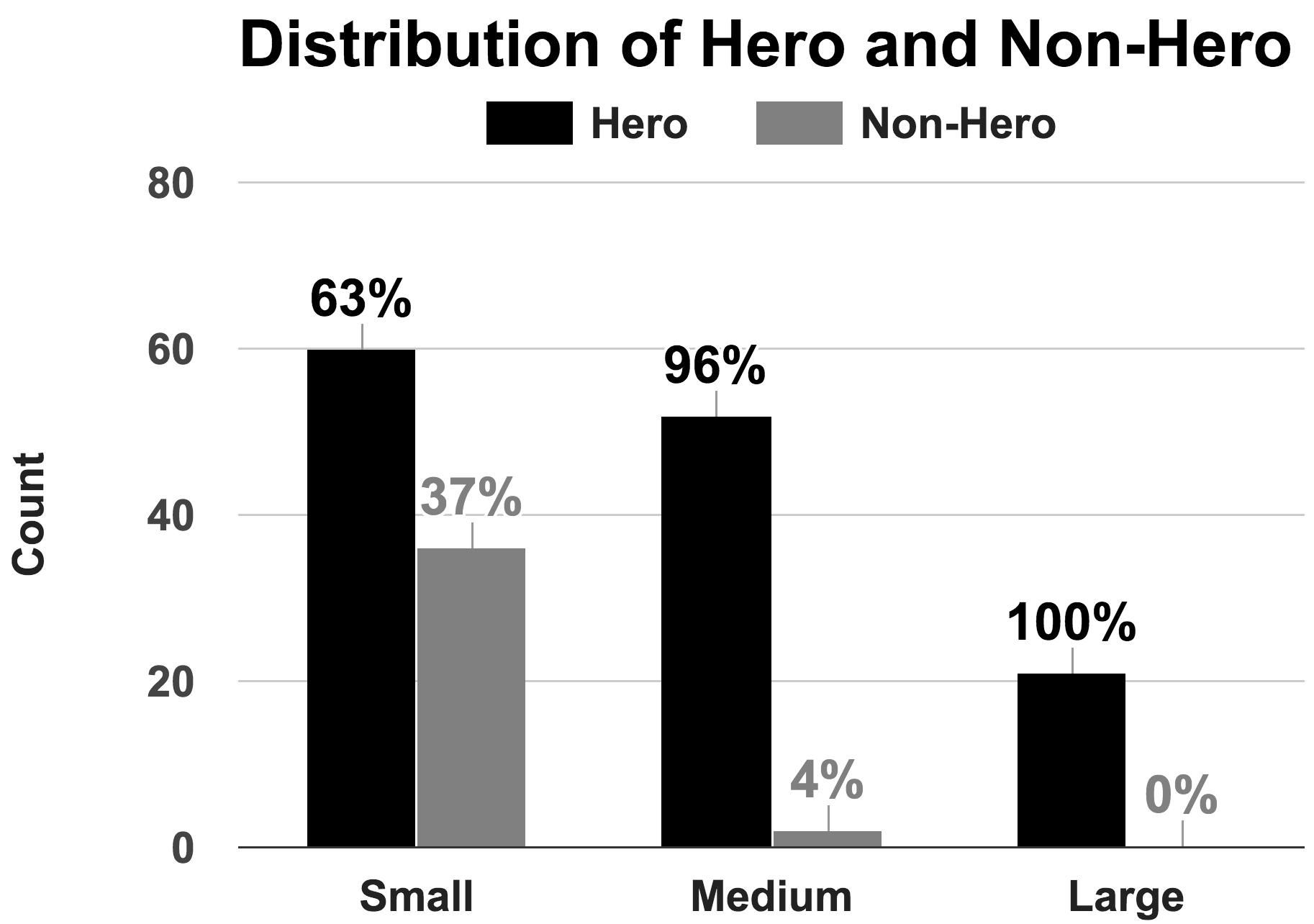}
  \caption{Enterprise projects:  Hero and Non Hero projects for different team sizes. As before,
  when the
  team size grows,    hero projects dominate our sample.}
  \label{fig:rq2b}
\end{figure*}

If the build effect does not explain these results, what does?
We think the  high frequency of heroes can be explained by the nature of software
development. For example, consider Github OSS projects,
they are often started by a project leader~\cite{ye2003toward} who is  responsible for maintaining and moderating that  project. 
Until the project becomes popular only the leader would be responsible to make major code contributions~\cite{tsay2014influence}. Once the project has become stable and popular, the on-going  issues/bugs/enhancement fixes are just few lines of code done by peripheral developers~\cite{pinto2016more}. Note that such a track record would naturally lead to heroes.

%
%

Whatever the reason, the pattern is very clear.
The ratio of hero projects in Figure~\ref{fig:rq1} is so large that it
motivates the rest of this paper. Accordingly, we move in to study the impact
of heroes on software quality.

\subsection{RQ2: How does team size affect the prevalence of hero projects?}

Figure~\ref{fig:rq2a} and \ref{fig:rq2b} show the distribution of Hero and non-hero projects across different team   sizes in Public and Enterprise respectively. The clear pattern in those results is
that as teams grow larger, they are more dependent on heroes. In fact, for large projects,
non-heroes almost disappear.

That is, contrary to established wisdom in the field~\cite{bier2011online},
what we  see here is most
projects make extensive use of heroes.
We conjecture that  the benefits of having heroes, where a small group handles the complex communications seen in large projects, out-weighs the theoretical drawbacks of heroes.




\begin{figure*}[!t]
\begin{minipage}{.33\linewidth}
\centering
         \includegraphics[width=\linewidth,keepaspectratio]{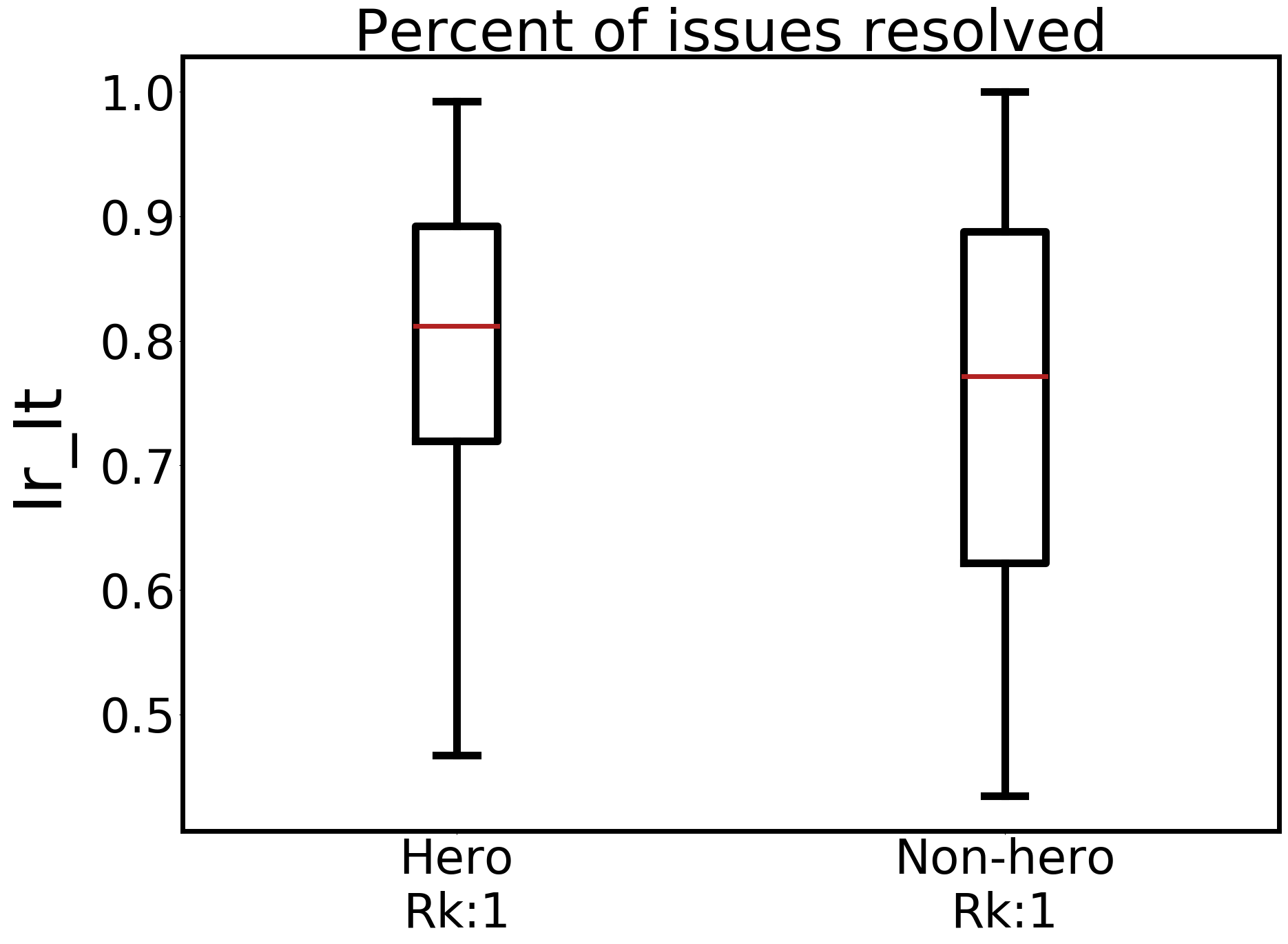}
    \end{minipage}%
\begin{minipage}{.33\linewidth}
        \centering
        \includegraphics[width=\linewidth]{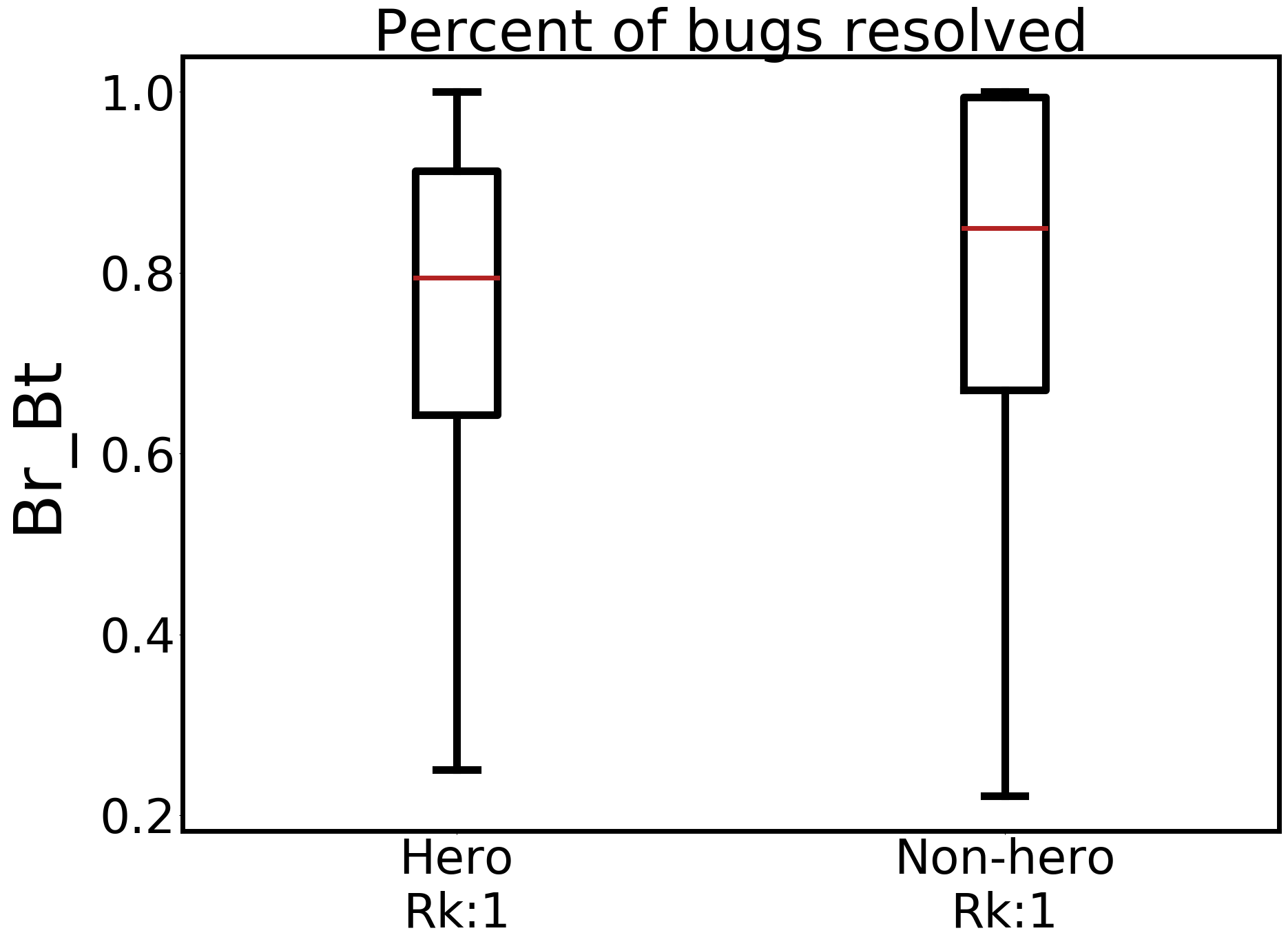}
    \end{minipage}%
\begin{minipage}{.33\linewidth}
        \centering
        \includegraphics[width=\linewidth]{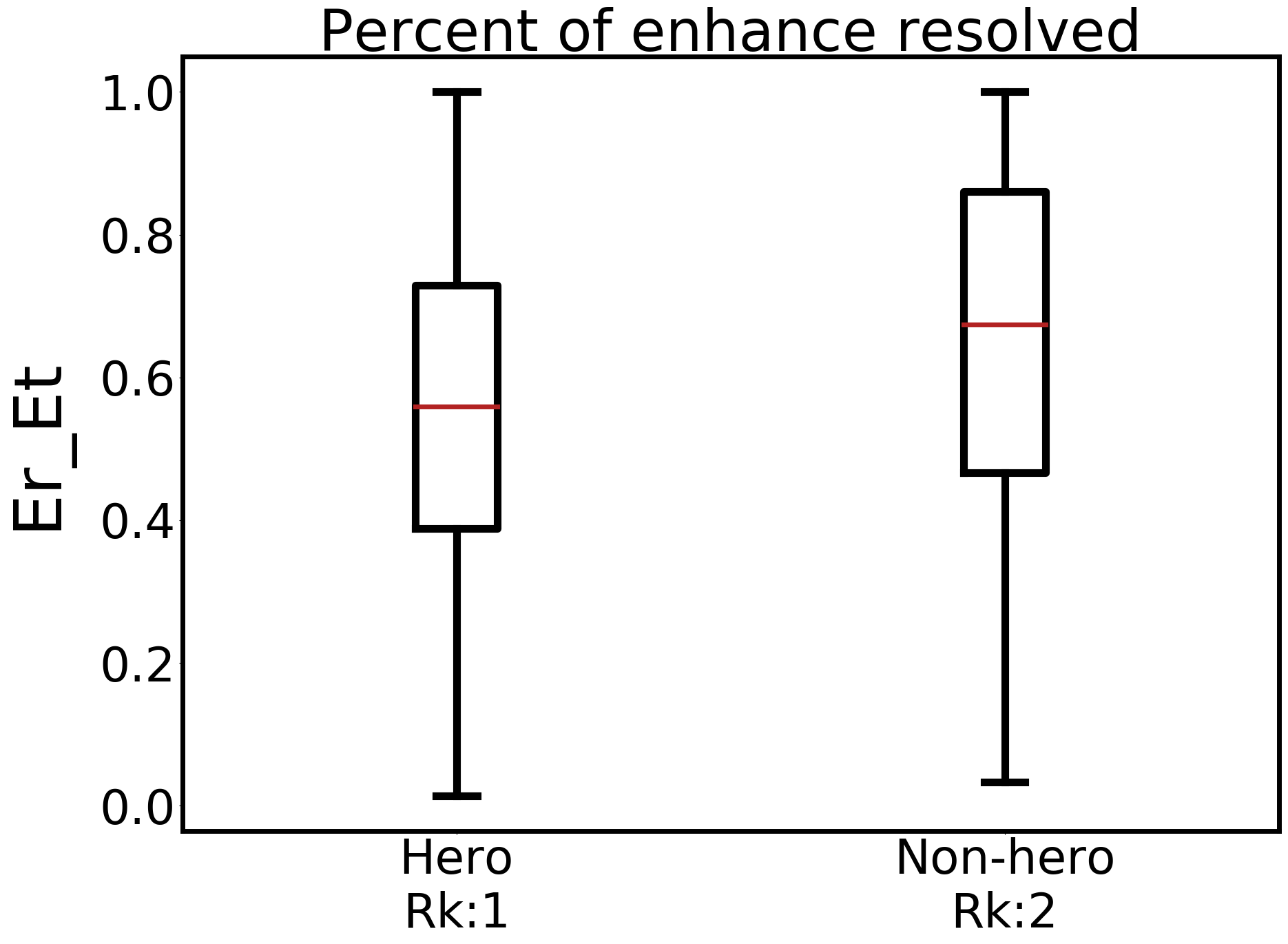}
    \end{minipage}%
    \caption{Public projects:  Hero and Non-hero values of $I_r\_I_t$, $B_r\_B_t$ and $E_r\_E_t$ (which is the ratio of issue, bug, enhancement reports being closed over the total issue, bug, enhancement reports created, respectively). Of these distributions, only the enhancement rates are different between hero and non-hero projects.}
    \label{fig:ratio_public}
\end{figure*}

\begin{figure*}[!t]
\begin{minipage}{.33\linewidth}
\centering
         \includegraphics[width=\linewidth,keepaspectratio]{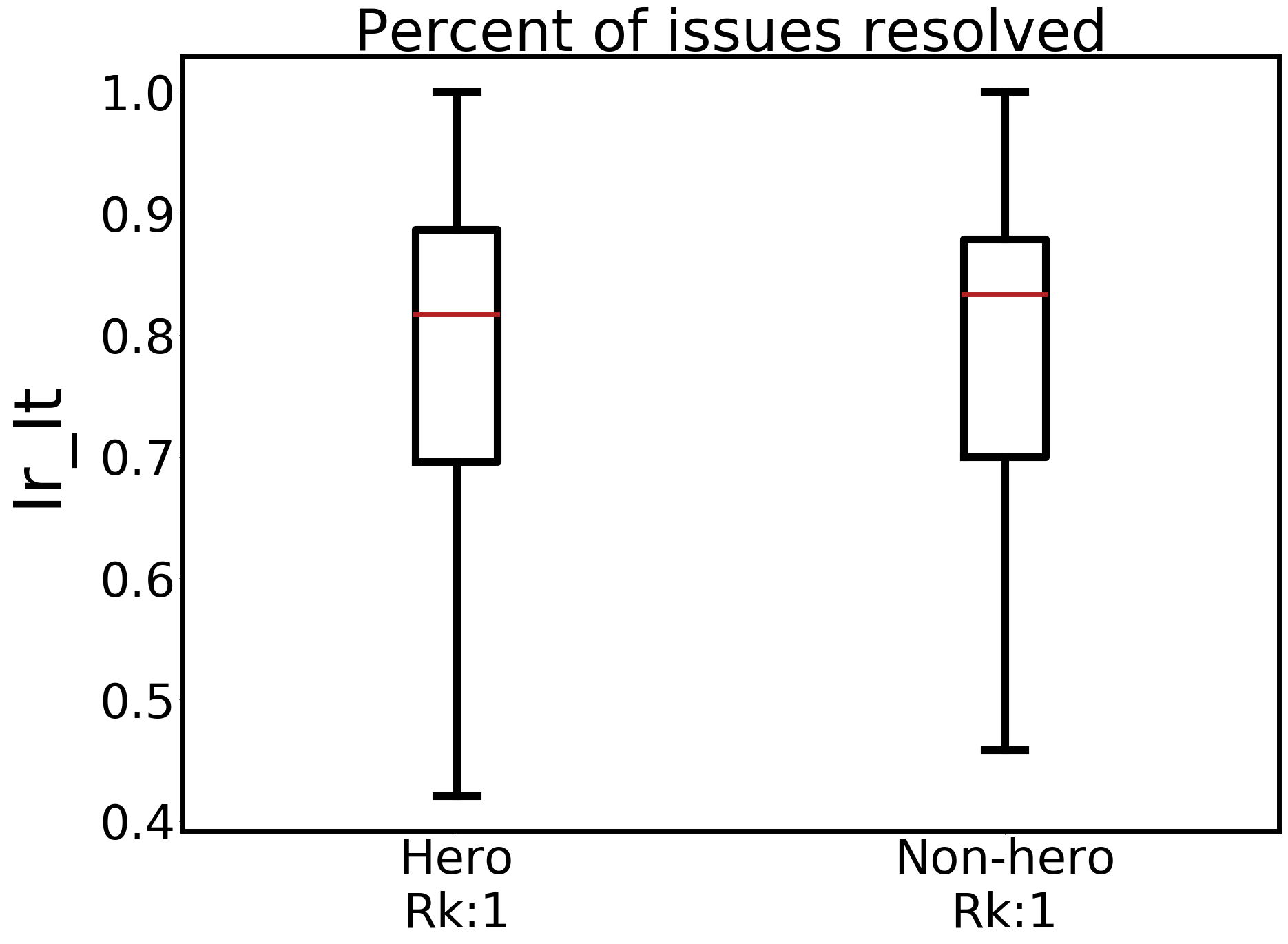}
    \end{minipage}%
\begin{minipage}{.33\linewidth}
        \centering
        \includegraphics[width=\linewidth]{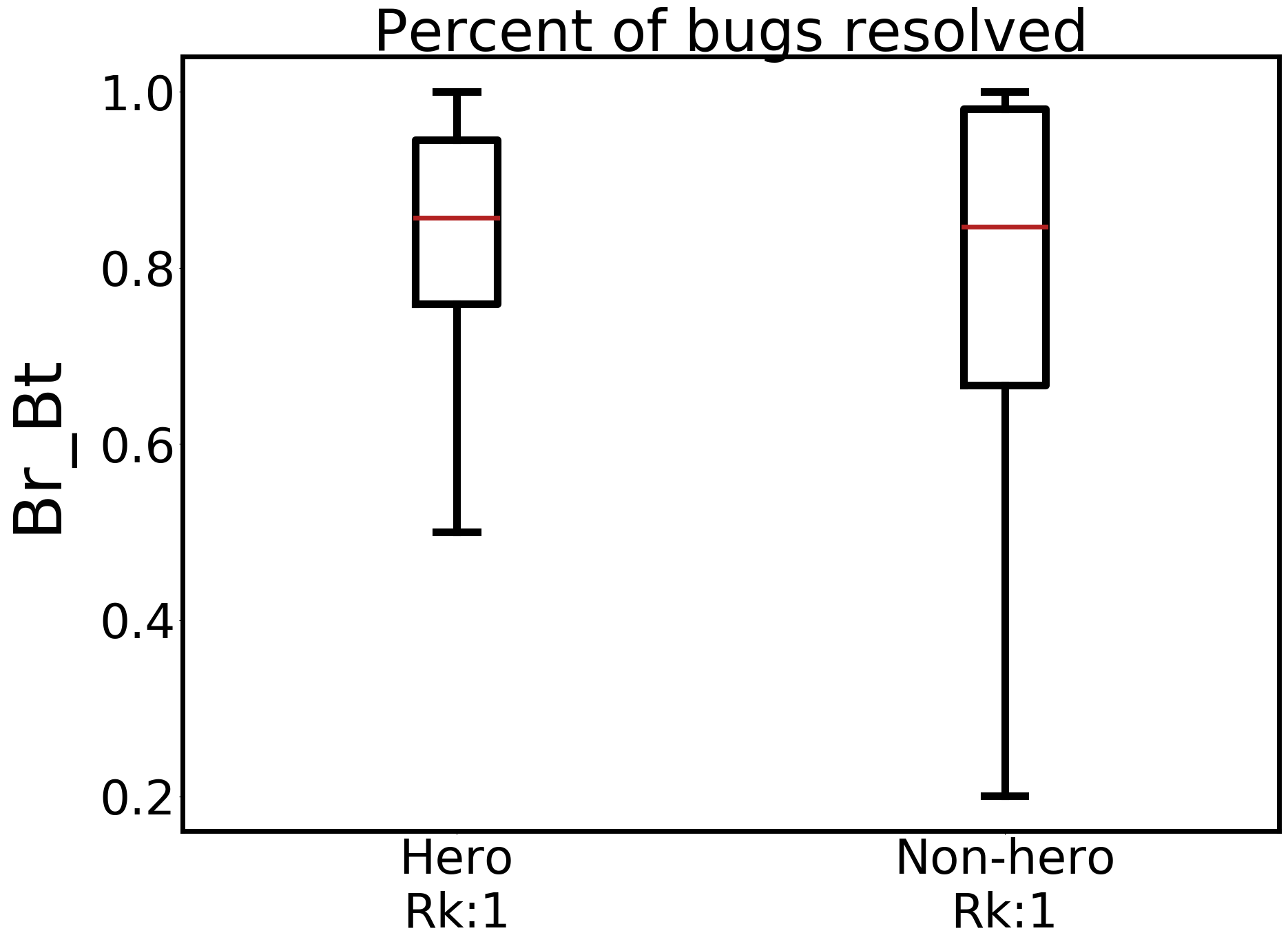}
    \end{minipage}%
\begin{minipage}{.33\linewidth}
        \centering
        \includegraphics[width=\linewidth]{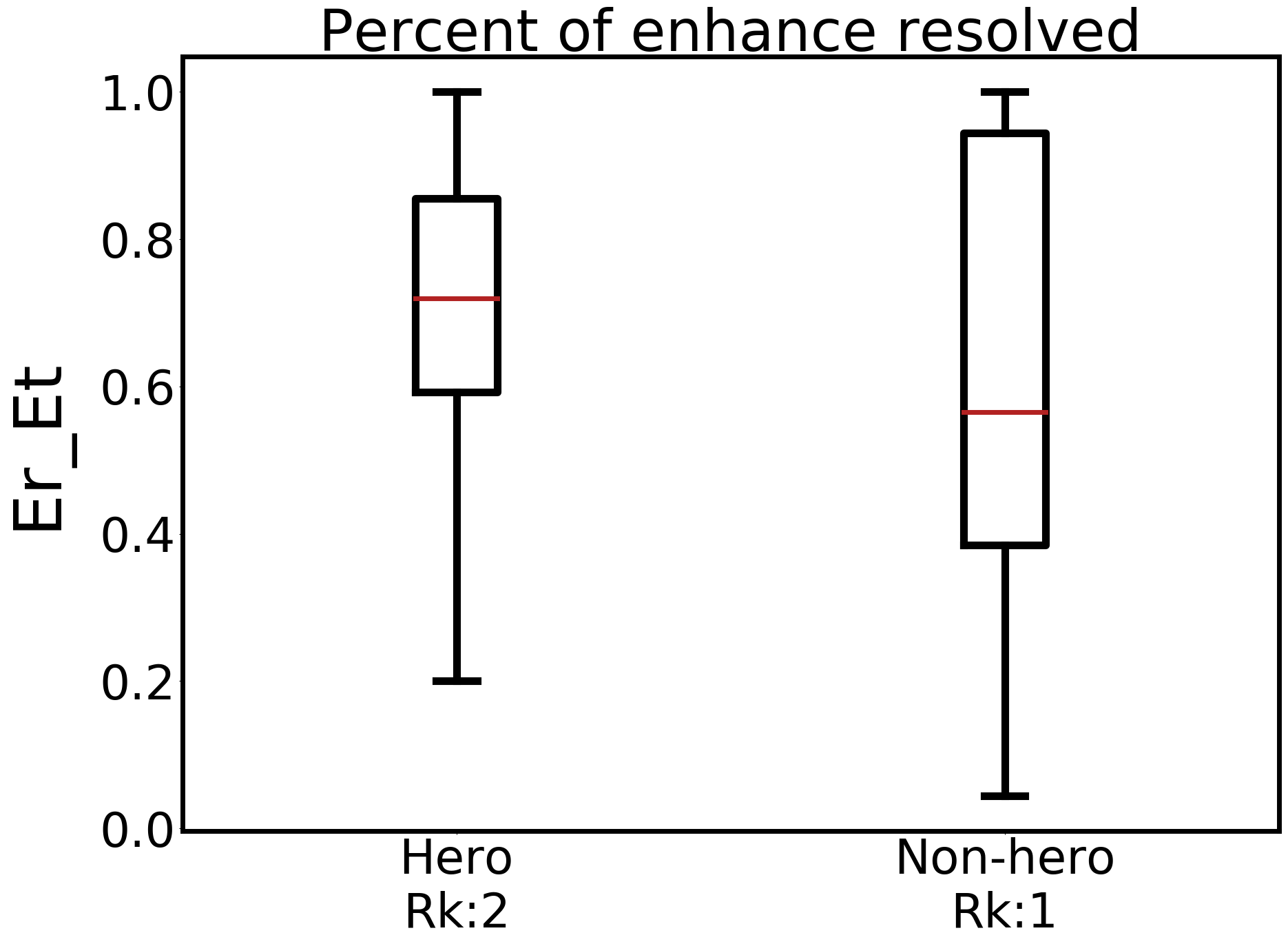}
    \end{minipage}%
    \caption{Enterprise projects:   Hero and Non-hero values of $I_r\_I_t$, $B_r\_B_t$ and $E_r\_E_t$ (which is the ratio of  issue, bug, enhancement reports being closed over the total issue, bug, enhancement reports created, respectively).
As before,  only the enhancement rates are different between hero and non-hero projects.
    }
    \label{fig:ratio_enter}
\end{figure*}

\subsection{RQ3: Are hero projects associated with  better software quality ?}
\label{sec:rq3}
We divide this investigation into two steps: RQ3a and RQ3b. RQ3a explores the {\em ratio} of issues/bugs/enhancements successfully closed.
Next, RQ3b explores the {\em time} required to close those issues.

\subsubsection{RQ3a: Does having a hero programmer improves the number of issues, bugs and enhancements being resolved?}
 
Figure~\ref{fig:ratio_public} and \ref{fig:ratio_enter} show   boxplots of each of   metrics
reporting the ratio of closed issue, bugs, and enhancements denoted by
$I_r\_I_t$, $B_r\_B_t$ and $E_r\_E_t$  respectively. Note that {\em larger} numbers are {\em better}.

In these figures,
the x-axis separates our Hero and Non-hero projects (found using the methods
of RQ1).  On the x-axis, each label is further labelled with ``Rk:1'' or ``Rk:2'' which is the result
of a statistical comparison of the two populations using the Scott-knott test explained in Section~\ref{sec:stats}. Note that in Figure~\ref{fig:ratio_public} and \ref{fig:ratio_enter}, for the issue and bug closed ratios,
the two distributions have the same rank, i.e., ``Rk:1''. This means that these populations
are statistically indistinguishable.

On the other hand, the ratio of closing enhancement issues in Public and Enterprise projects
is statistically distinguishable, as shown by the ``Rk:1'' and ``Rk:2'' labels on those plots.
Interestingly, the direction of change is different in Public and Enterprise projects:
\bi
\item
In Public projects, heroes close the {\em fewest} enhancement issues;
\item
But in Enterprise projects, heroes close the {\em most} enhancement issues;
\item
Further, in Enterprise projects, the variance in the percentage of closed enhancements is much
smaller with heroes than otherwise. That is, heroes in Enterprise development result
in more control of that project.
\ei
Hence, while we should   depreciate hero projects for open source projects,
we should encourage them for Enterprise projects. Note that this is very much
the opposite of conventional wisdom~\cite{bier2011online}. That said,
our reading of the literature is that heroes have been studied much more in OSS
projects than in proprietary Enterprise projects. Hence, this finding (that proprietary Enterprise projects benefit from heroes) might have existed undetected for some time.

\begin{figure*}[!t]
\begin{minipage}{.33\linewidth}
\centering
         \includegraphics[width=\linewidth,keepaspectratio]{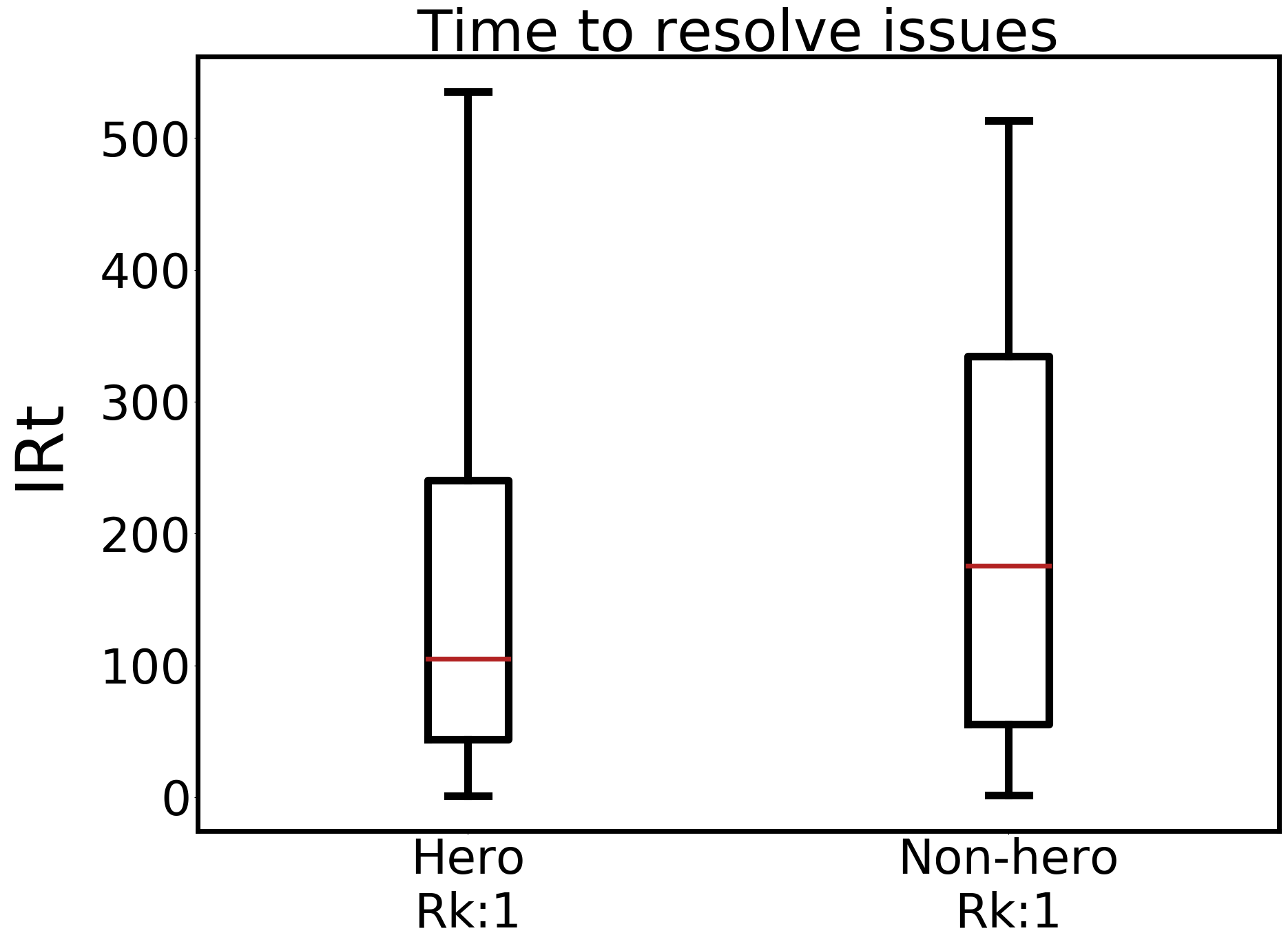}
    \end{minipage}%
\begin{minipage}{.33\linewidth}
        \centering
        \includegraphics[width=\linewidth]{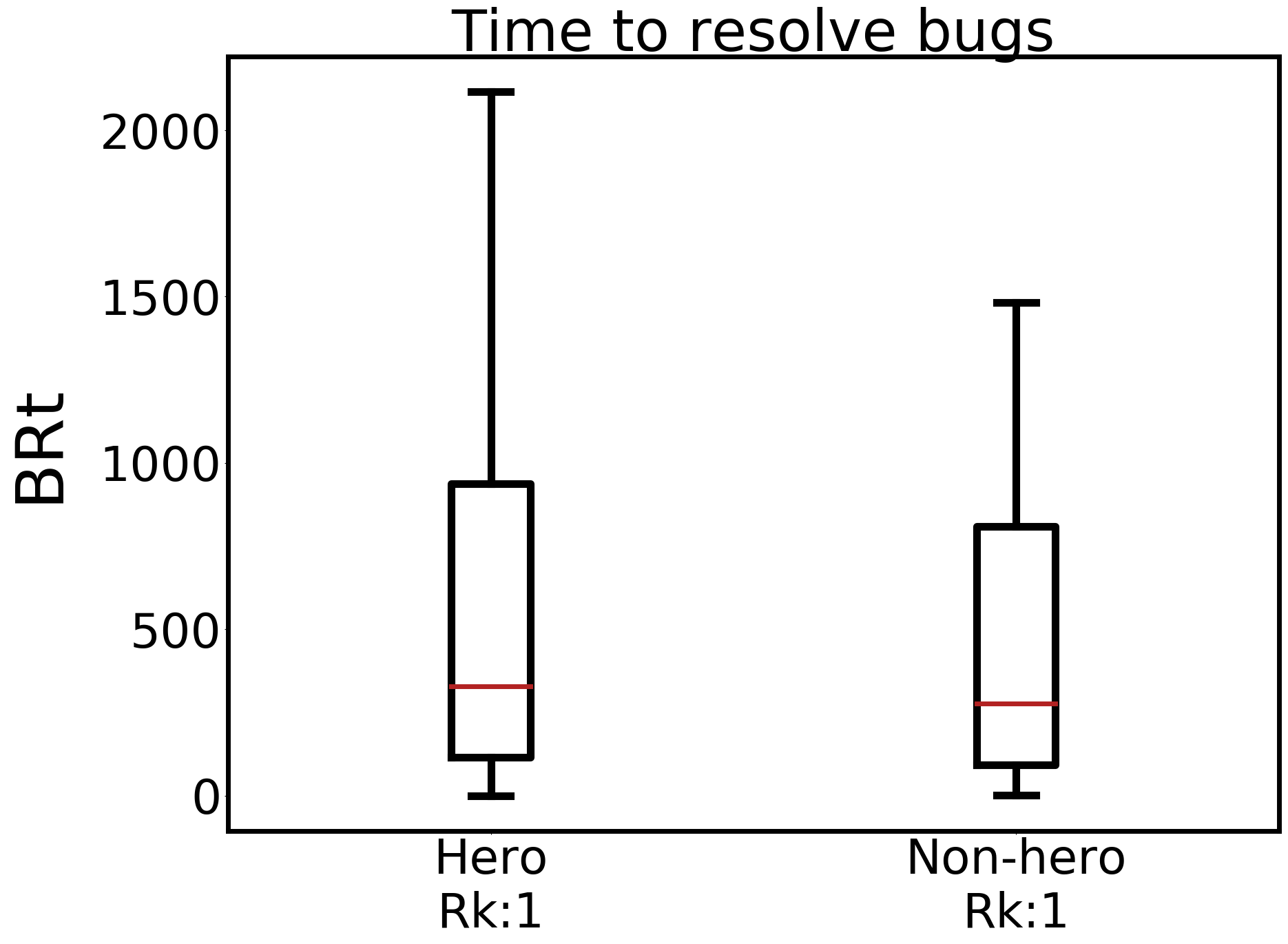}
    \end{minipage}%
\begin{minipage}{.33\linewidth}
        \centering
        \includegraphics[width=\linewidth]{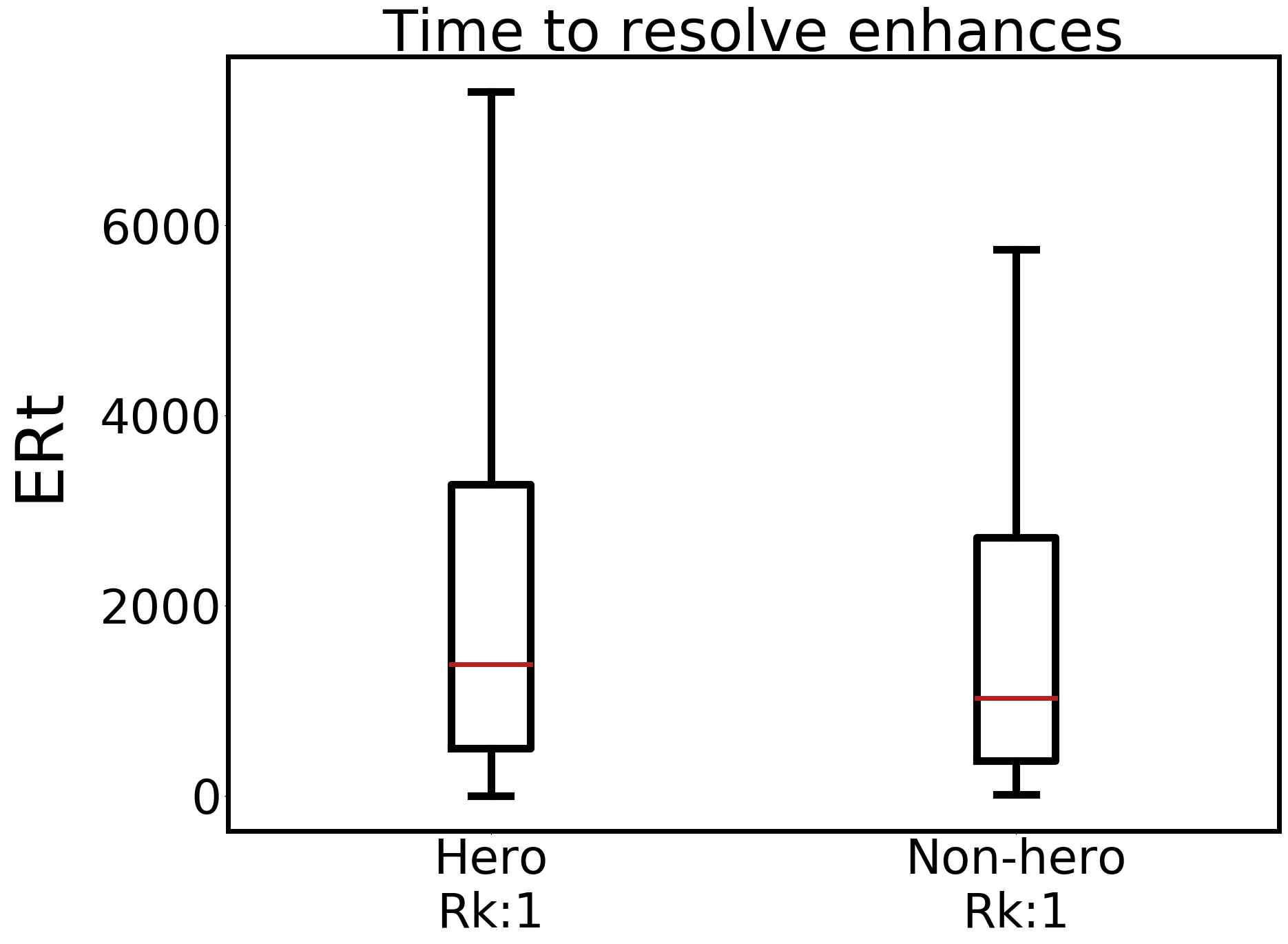}
    \end{minipage}%
    \caption{Public projects: Hero and Non-hero values  of  $IR_t$, $BR_t$ and $ER_t$ (which is the median time taken to resolve issue, bugs, enhancement reports respectively). Y-axis shown is in hours.}
    \label{fig:resolution_public}
\end{figure*}

\begin{figure*}[!t]
\begin{minipage}{.33\linewidth}
\centering
         \includegraphics[width=\linewidth,keepaspectratio]{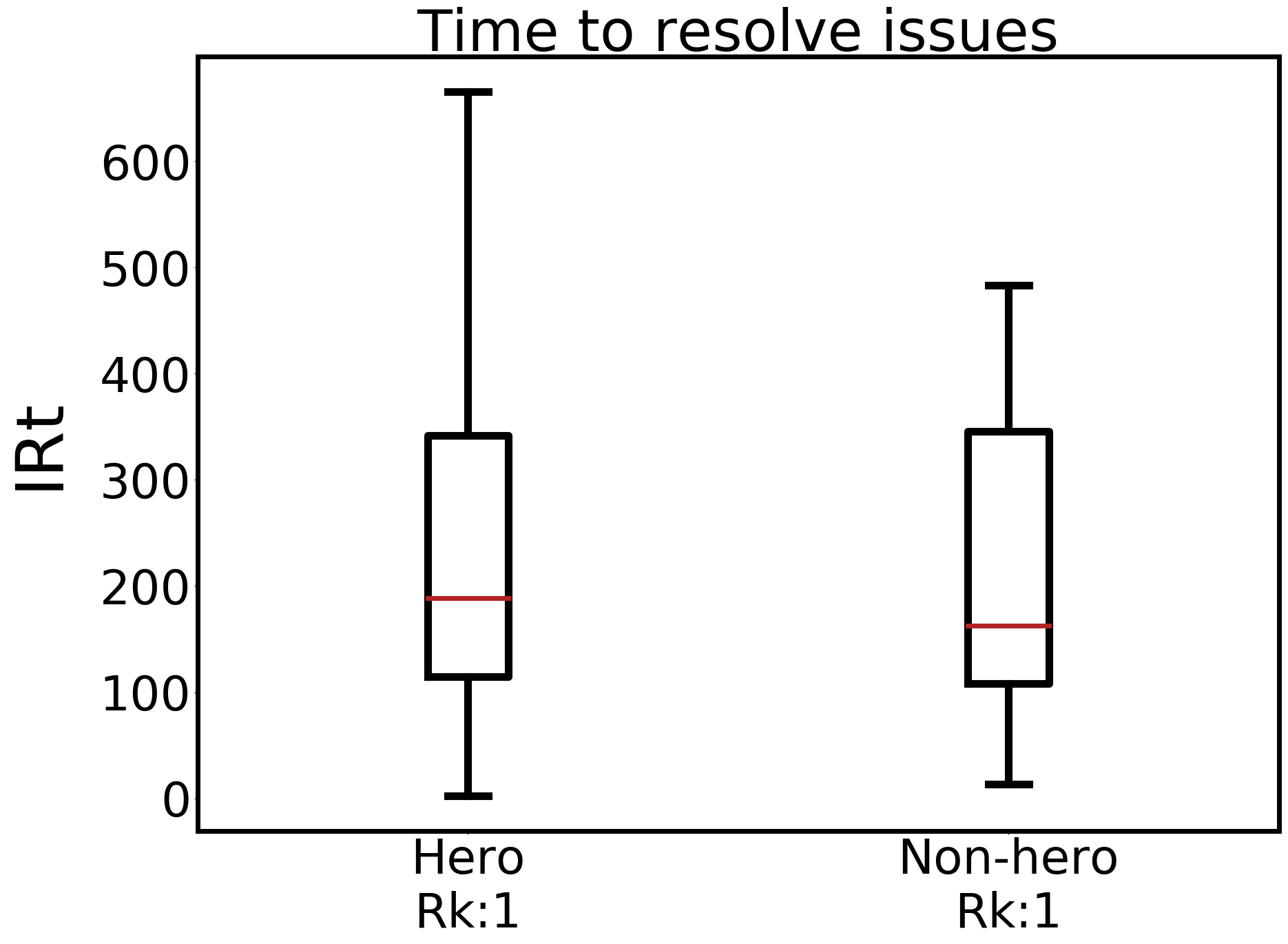}
    \end{minipage}%
\begin{minipage}{.33\linewidth}
        \centering
        \includegraphics[width=\linewidth]{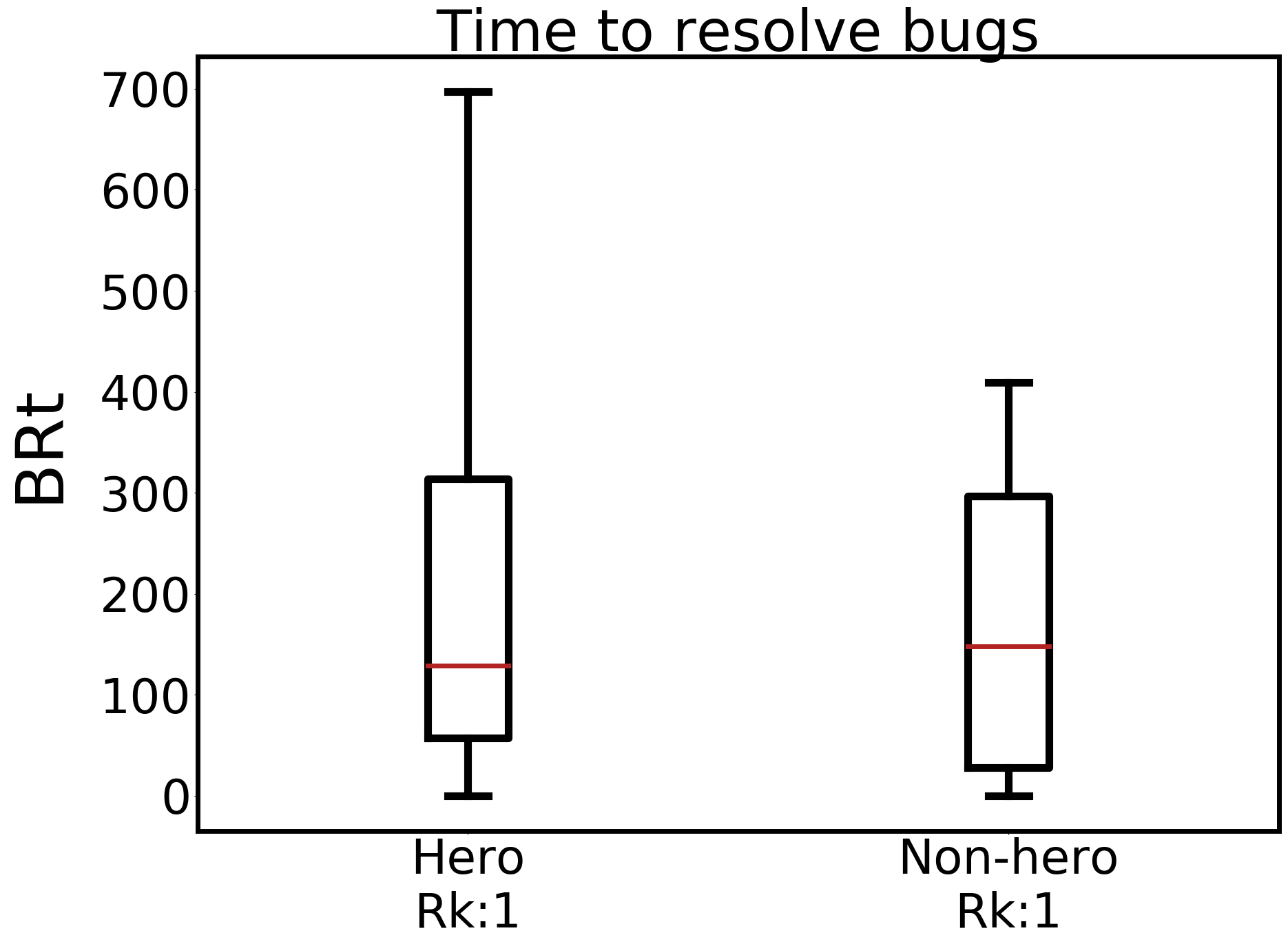}
    \end{minipage}%
\begin{minipage}{.33\linewidth}
        \centering
        \includegraphics[width=\linewidth]{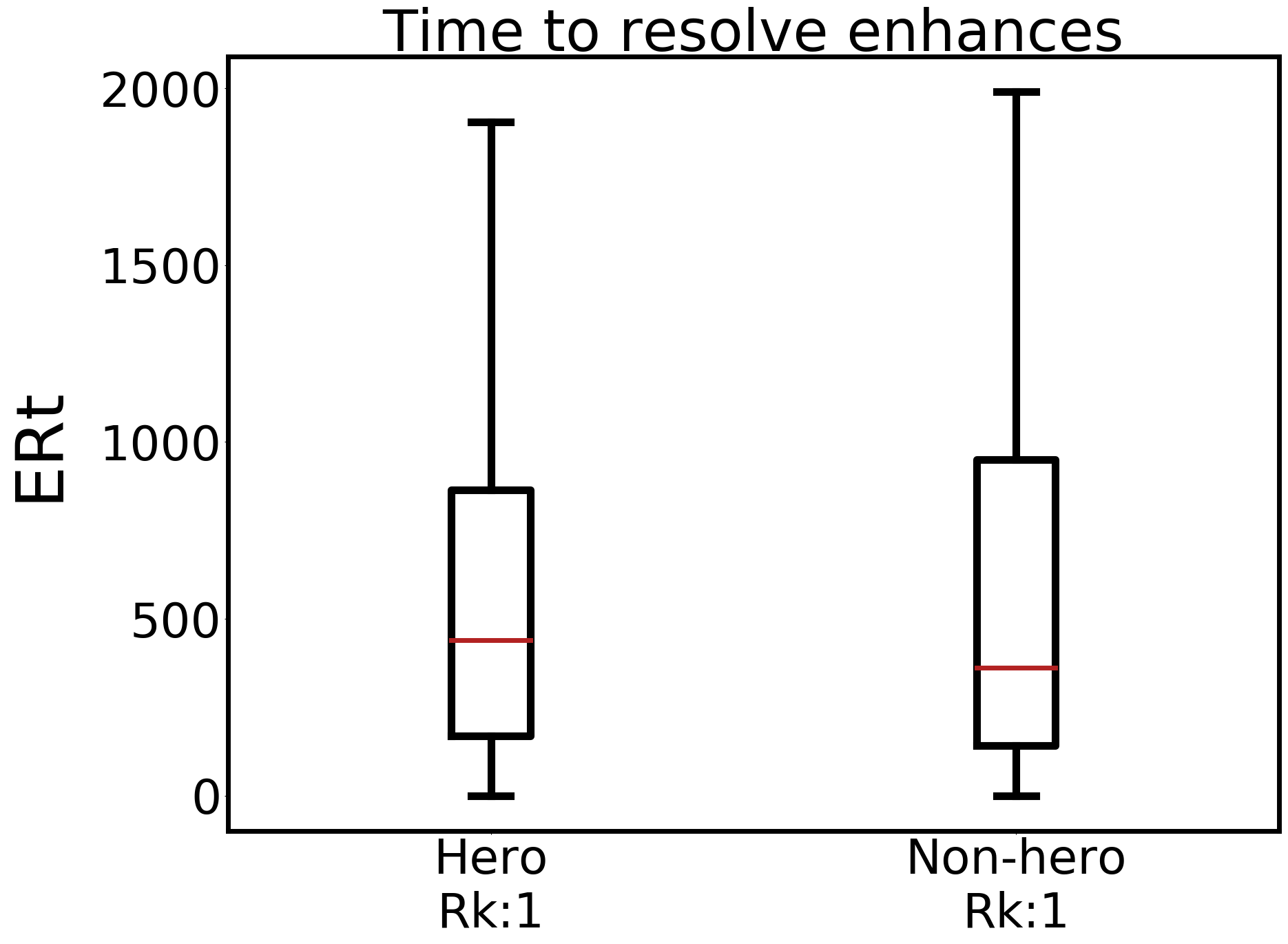}
    \end{minipage}%
    \caption{Enterprise projects:   Hero and Non-hero values of  $IR_t$, $BR_t$ and $ER_t$ (which is the median time taken to resolve issue, bugs, enhancement reports respectively). Y-axis shown is in hours.}
    \label{fig:resolution_enterprise}
\end{figure*}

\subsubsection{RQ3b: Does having a hero programmer improves the time to resolve issues, bugs and enhancements?}

Figure~\ref{fig:resolution_public} and \ref{fig:resolution_enterprise} show   boxplots of   reporting
the time required to close issues, bugs, and enhancements denoted  by $IR_t$, $BR_t$ and $ER_t$ respectively.   Note that for these figures, {\em smaller} numbers are {\em better}.

Like before, the x-labels are marked with the results of a statistical comparison of
these pairs of distributions. Note that all these statistical ranks are  ``Rk:1'', i.e., 
all these pairs of distributions are statistically indistinguishable.
That is, there is no effect to report here about effect of heroes or non-heroes on
the time required to close issues, bugs and enhancements.

\section{Discussion}
\label{sec:discuss}
What's old is new. Our results (that heroes
are important) echo a decades old concept.
In 1975, Fred Brooks wrote of  ``surgical teams'' and the ``chief programmer''~\cite{brooks1975mythical}.
He argued that:
\bi
\item
Much as a surgical team during surgery is led by one surgeon performing the most critical work, while directing the team to assist with less critical parts,
\item
Similarly, software projects should be led by one ``chief programmer'' to  develop critical system components while the rest of a team provides what is needed at the right time. 
\ei
 Brooks conjecture that ``good'' programmers are generally five to ten times as productive as mediocre ones. We note that our definition of ``hereos'' (80\% of the work done by 20\% of the developers) is consistent with the Brooks's conjecture that heroes are  five times more productive than the other team members.
 
 Prior to this research, we had thought that in the era of open source and agile, all such notions of ``chief programmers'' and ``heroes'' were historical relics, and that development teams would now be distributing the workload across the whole project. 
 
 But based on the results of this paper, we have a different view. Projects are written by people of various levels of skills. Some of those people are so skilled that they become the project heroes. Organizations need to acknowledge their dependency on such heroes, perhaps altering their human resource policies. Specifically,  organizations need to recruit and retain more heroes (perhaps by offering heroes larger annual bonuses).



\section{Threats to Validity}
\label{sec:validity}

As with any large scale empirical study, biases can affect the final
results. Therefore, any conclusions made from this work
must be considered with the following issues in mind:

\bi 

\item \textit{Internal Validity}
 
    \bi
    \item \textit{Sampling Bias}: Our conclusions are based on the 1,108+538 Public+Enterprise Github projects
    that started this analysis.  It is possible that   different initial projects would have lead to different conclusions. That said, our initial sample is very large so we have some
    confidence that this sample
    represents an interesting range of projects. As evidence of that, we note that our sampling bias is less pronounced than other Github studies since we explored {\em both} Public and Enterprise projects (and many prior studies only explored Public projects.
    \item \textit{Evaluation Bias}: 
    In  RQ3b, we said that there is no difference between heroes or non-heroes on
the time required to close issues, bugs and enhancements. While that statement is true, that conclusion is scoped by the evaluation metrics we used to write this paper. It is possible that, using other measurements, there may well be a difference in these different kinds of projects. This is a matter that needs to be explored in future research. 
    \ei
    
\item \textit{Construct Validity}: At various places in this report, we made engineering decisions about (e.g.) team size and what
constitutes a ``hero'' project. While
those decisions were made using advice from
the literature (e.g.~\cite{gautam2017empirical}),
we acknowledge that other constructs might lead to different conclusions. 

\item \textit{External Validity}:  We have relied on issues marked as a `bug' or `enhancement' to count bugs or enhancements, and bug or enhancement resolution times. In Github, a bug or enhancement might not be marked in an issue but in commits. There is also a possibility that the team of that project might be using different tag identifiers for bugs and enhancements. To reduce the impact of this problem, we  did take precautionary step to (e.g.,) include various tag identifiers from Cabot et al.~\cite{cabot2015exploring}. We also took precaution to remove any pull merge requests from the commits to remove any extra contributions added to the hero programmer. 

\item \textit{Statistical Validity}: To increase
the validity of our results, we applied
 two statistical tests, bootstrap and the a12.
 Hence, anytime in this paper we reported that ``X was different from Y'' then that report
 was based on both an effect size
 and a statistical significance test.
\ei

\section{Conclusion}
\label{sec:concl}

The established wisdom in the literature is to depreciate ``heroes'',
i.e., a small percentage of the staff
responsible for most of the progress on a project.
After mining 661 Public and 171 Enterprise Github projects,
we assert that it is  time to revise that wisdom:
\bi
\item Overwhelmingly, most projects are hero projects, particularly when we look at medium to large projects. That is, discussions about the merits of avoiding heroes is really relevant only to smaller projects.
\item
Heroes do not significantly affect the rate at which issues or bugs are closed.
\item 
Nor do they influence the time required to address issues, bugs or enhancements.
\item
Heroes positively influence the rate at which enhancement requests are managed
within Enterprise project.
\ei
The only place where our results agree with established wisdom is for the enhancement rates
for non-hero Public projects. In this particular case, we saw that non-hero projects
are enhanced fastest. That said, given the first point listed above,  that benefit
for non-hero projects is very rare.

In summary, our empirical results call for a revision of a long-held truism in software engineering.
Software heroes are far more common and valuable than suggested by the literature,
particularly for medium to large  Enterprise developments. Organizations should reflect on better ways
to find and retain more of these heroes.

\section{Acknowledgements}
\label{sec:ack}

The first and second authors conducted this research study as part
of their internship at the industry in Summer, 2017. We also express our gratitude to our industrial  partner for providing us the opportunity to mine hundreds
of their
Enterprise projects. Also,
special thanks to our colleagues and mentors at the industry for
their valuable feedback.

\balance

\bibliographystyle{ACM-Reference-Format}


\end{document}